\documentclass[a4paper]{spie}  

\usepackage[]{graphicx}
\usepackage{amssymb,amsmath,psfrag,url,color}

\newcommand{\grad}{\ensuremath{^\circ}}

\newcommand{\MyField}[1]{{\bf{#1}}}

\newcommand{\MyTensor}[1]{{{{#1}}}}

\title{Metrology of EUV Masks by EUV-Scatterometry and Finite Element Analysis} 

\author{
Jan Pomplun\supit{\,ab},
Sven Burger\supit{\,ab}, 
Frank Schmidt\supit{\,ab}, \\
Frank Scholze\supit{\,c},
Christian Laubis\supit{\,c},
Uwe Dersch\supit{\,d},
\skiplinehalf
\supit{a}
Zuse Institute Berlin,
Takustra{\ss}e 7,
D\,--\,14\,195 Berlin,
Germany
\smallskip\\
\supit{b}
JCMwave GmbH,
Haarer Stra{\ss}e 14a,
D\,--\,85\,640 Putzbrunn, 
Germany
\smallskip\\
\supit{c}
Physikalisch-Technische Bundesanstalt,
EUV radiometry,
Abbestra{\ss}e 2\,--\,12,
D\,--\,10\,587 Berlin,
Germany
\smallskip\\
\supit{d}
Advanced Mask Technology Center GmbH \& Co. KG,
R{\"a}hnitzer Allee 9,
D\,--\,01\,109 Dresden,
Germany
}

\authorinfo{
Corresponding author: J. Pomplun\\
URL: http://www.zib.de/nano-optics/\\
Email: pomplun@zib.de
}

\begin{document} 
  \maketitle 


\noindent
This paper has been published in Proc.~SPIE Vol. {\bf  7028}
(2008) 70280P, 
({\it Photomask and Next-Generation Lithography Mask Technology XV, Toshiyuki Horiuchi, Editor})
and is made available as an electronic preprint with permission of SPIE. 
Copyright 2008 Society of Photo-Optical Instrumentation Engineers. 
One print or electronic copy may be made for personal use only. 
Systematic reproduction and distribution, duplication of any material in this paper for a fee or for 
commercial purposes, or modification of the content of the paper are prohibited. 

\begin{abstract}
Extreme ultraviolet (EUV) lithography is seen as a main candidate for production of future generation computer technology. Due to the short wavelength of EUV light ($\approx 13\,$nm) novel reflective masks have to be used in the production process. A prerequisite to meet the high quality requirements for these EUV masks is a simple and accurate method for absorber pattern profile characterization.

In our previous work we demonstrated that the Finite Element Method (FEM) is very well suited for the simulation of EUV scatterometry and can be used to reconstruct EUV mask profiles from experimental scatterometric data \cite{Pomplun2007,Scholze07}. 

In this contribution we apply an indirect metrology method to periodic EUV line masks with different critical dimensions ($140\,$nm and $\,540\,$nm) over a large range of duty cycles (1:2, ... , 1:20). We quantitatively compare the reconstructed absorber pattern parameters to values obtained from direct AFM and CD-SEM measurements. We analyze the reliability of the reconstruction for the given experimental data. For the CD of the absorber lines, the comparison shows agreement of the order of 1nm.

Furthermore we discuss special numerical techniques like domain decomposition algorithms and high order finite elements and their importance for fast and accurate solution of the inverse problem.
\end{abstract}


\section{Introduction}
The quality of pattern profiles of EUV masks becomes important e.g. due to shadowing effects at oblique incidence illumination \cite{Sugawara05a,Sugawara05b}. Consequently, there is a need for adequate destruction free pattern profile metrology techniques, allowing characterization of mask features down to a typical size of $100\,$nm and below. The desired accuracy is thereby of the order of $1\,$nm which is challenging even for direct atomic force (AFM) or scanning electron microscopic (SEM) measurements. These methods e.g. suffer from surface charges and the need of the definition of edge operators to reconstruct the CD. Furthermore it is extremely time consuming and not practical to scan large areas of a mask with direct microscopical measurements.

Here the usage of scatterometry as a metrology technique offers the possibility to analyze larger areas on a mask depending on the diameter of the incident beam, e.g. arrays of periodic absorber lines, contact holes etc. Of course detecting small errors of a single feature remains extremely challenging.
\section{Characterization of EUV masks by EUV scatterometry}
\label{sec::scatterometry}
\begin{figure}
(a)\hspace{9cm}(b)\\\vspace{0.4cm}\\
\psfrag{euv}{light source}
\psfrag{mult}{Mo/Si multilayer}
\psfrag{det}{detector with slit}
\psfrag{abs}{absorber stack}
\psfrag{Si}{Si-cap}
\psfrag{cd}{CD}
\includegraphics[height=5cm]{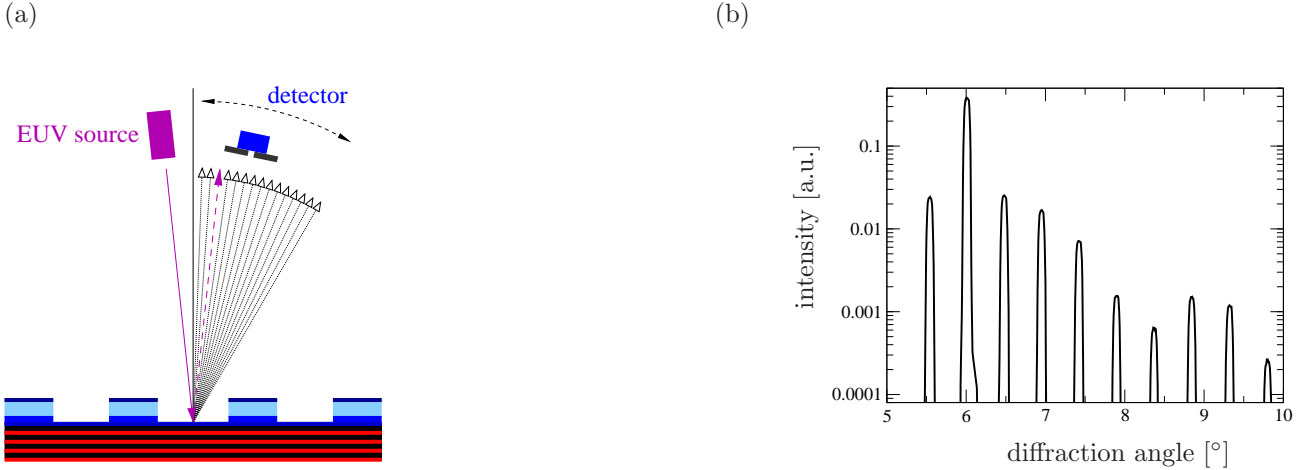}
\hfill
\psfrag{diff}{diffraction angle [\grad]}
\psfrag{intensity}{intensity [a.u.]}
\includegraphics[height=5cm]{fig/h4scatter.eps}
\caption{\label{fig:h4scatter} (a) Experimental setup for scatterometry experiment with fixed incident angle of $\theta_{in}=6\grad$ and variable angle of detection $\theta_{out}$; (b) Result of single wavelength scatterometry measurement at $\lambda=13.655\,$nm. Diffraction orders appear as peaks with finite width, the zeroth diffraction peak is centered around $6\grad$.}
\end{figure}

Single wavelength scatterometry, the analysis of light diffracted from a periodic structure, is a well suited tool for analysis of the geometry of EUV masks. Since scatterometry only needs a light source and a simple detector with no imaging lens system, its setup is inexpensive and offers no additional technical challenges. Fig. \ref{fig:h4scatter}(a) shows a sketch of the experimental setup. Light of fixed wavelength and fixed incident angle is reflected from the mask and the intensity of the reflected light is measured in dependence on the diffraction angle.
The use of EUV light for mask characterization is advantageous because it fits the small feature sizes on EUV masks. Diffraction phenomena are minimized, and of course the appropriate wavelength of the resonant structure of the underlying multilayer is chosen. Light is not only reflected at the top interface of the mask but all layers in the stack contribute to reflection. Therefore one expects that EUV radiation provides much more information on relevant EUV mask features than conventional long wavelength methods. The presented EUV measurements often provide up to 20 or more non-evanescent diffraction orders.

All measurements for the present work were performed by the Physikalisch-Technische Bundesanstalt (PTB) at the electron storage ring BESSY II \cite{Ulm98}. PTB's EUV reflectometer installed at the soft X-ray radiometry beamline allows high-accuracy measurements of very large samples with diameters up to $550\,$mm \cite{Scholze05,Scholze03}.

Experimental data was taken from an EUV test mask fabricated by AMTC. The mask is subdivided into 11 (labeled A-K) times 11 (labeled 1-11) test fields. Each of these test fields furthermore consists of a number of differently patterned areas allowing e.g. bright field, dark field and CD uniformity measurements. In the present work fields with periodic absorber lines of varying pitch and duty cycle are considered. We focus on fields D4, H4, F6, D8, H8 (uniformity fields with $720\,$nm pitch and $540\,$nm CD) and fields H1 to H5 (varying pitch and $140\,$nm CD), see Table \ref{table:mask}.

\begin{table}[h]
\begin{centering}
\begin{tabular}{lc}
stack & test mask \\
\hline
ARC + TaN-absorber& $67\,$nm\\
 SiO$_{2}$-buffer & $10\,$nm\\
Si-capping layer& $11\,$nm\\
multilayer & Mo/Si \\
\end{tabular}
\hspace{3cm}
\begin{tabular}{lcccc}
test field &CD [nm]& pitch [nm]& duty cycle\\
\hline
H1 & 140 &  2940 & $1:20$\\
H2 & 140 &  2240 & $1:15$\\
H3 & 140 &  1540 & $1:10$\\
H4 & 140 &  840 & $1:5$\\
H5 & 140 &  520 & $1:2$\\
D4, H4, F6, D8, H8& 540 &720 &$3:1$
\end{tabular}
\end{centering}
\caption{\label{table:mask}Design parameters (see also Fig. \ref{fig:defineParam}) for EUV test mask and test fields produced by AMTC.}
\end{table}
\section{FEM simulation of EUV scatterometry}
\label{sec::fem}
\begin{figure}[t]
\includegraphics[width=4cm,height=2.63cm]{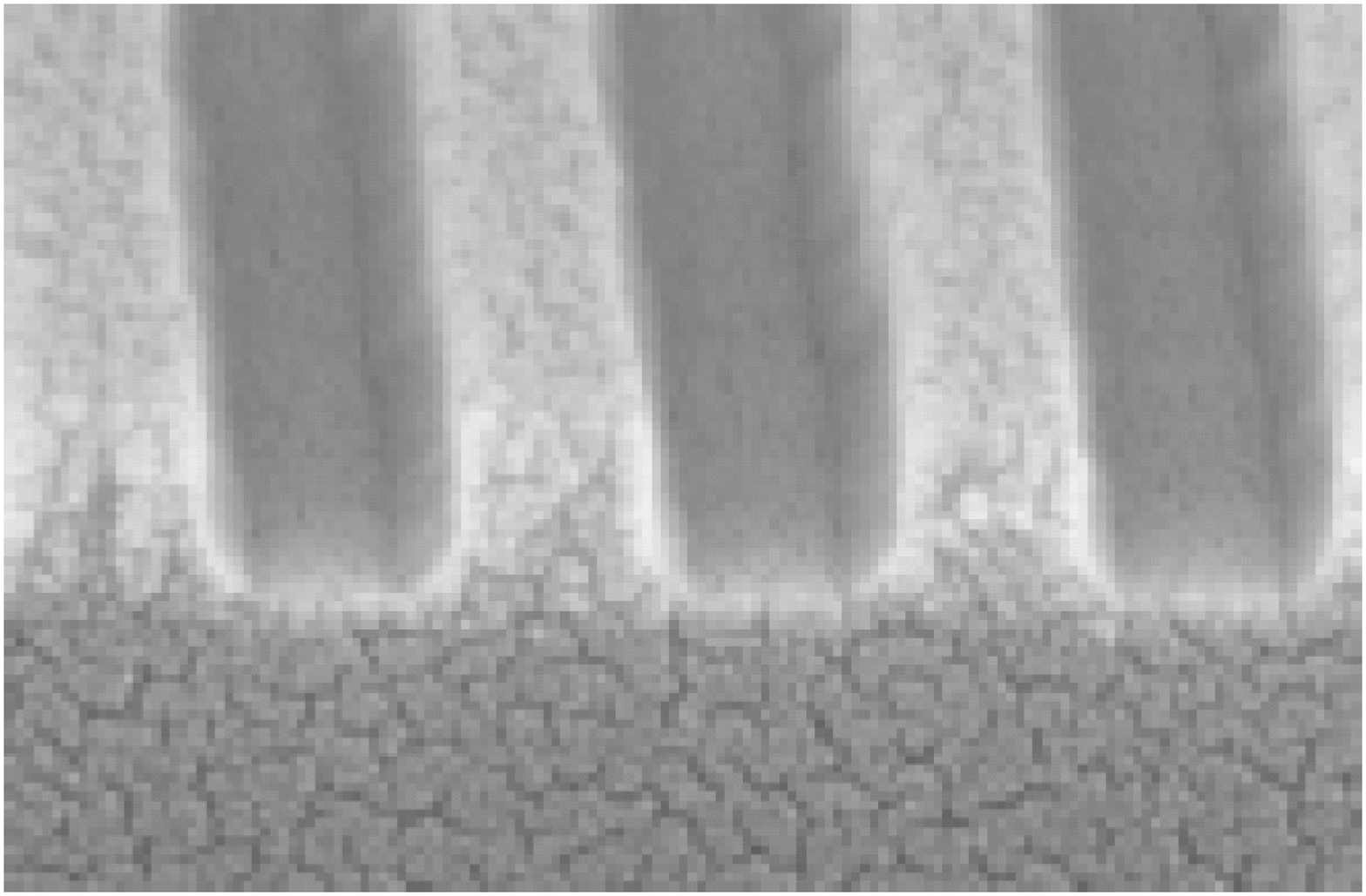}\hfill
\includegraphics[width=4cm,height=2.63cm]{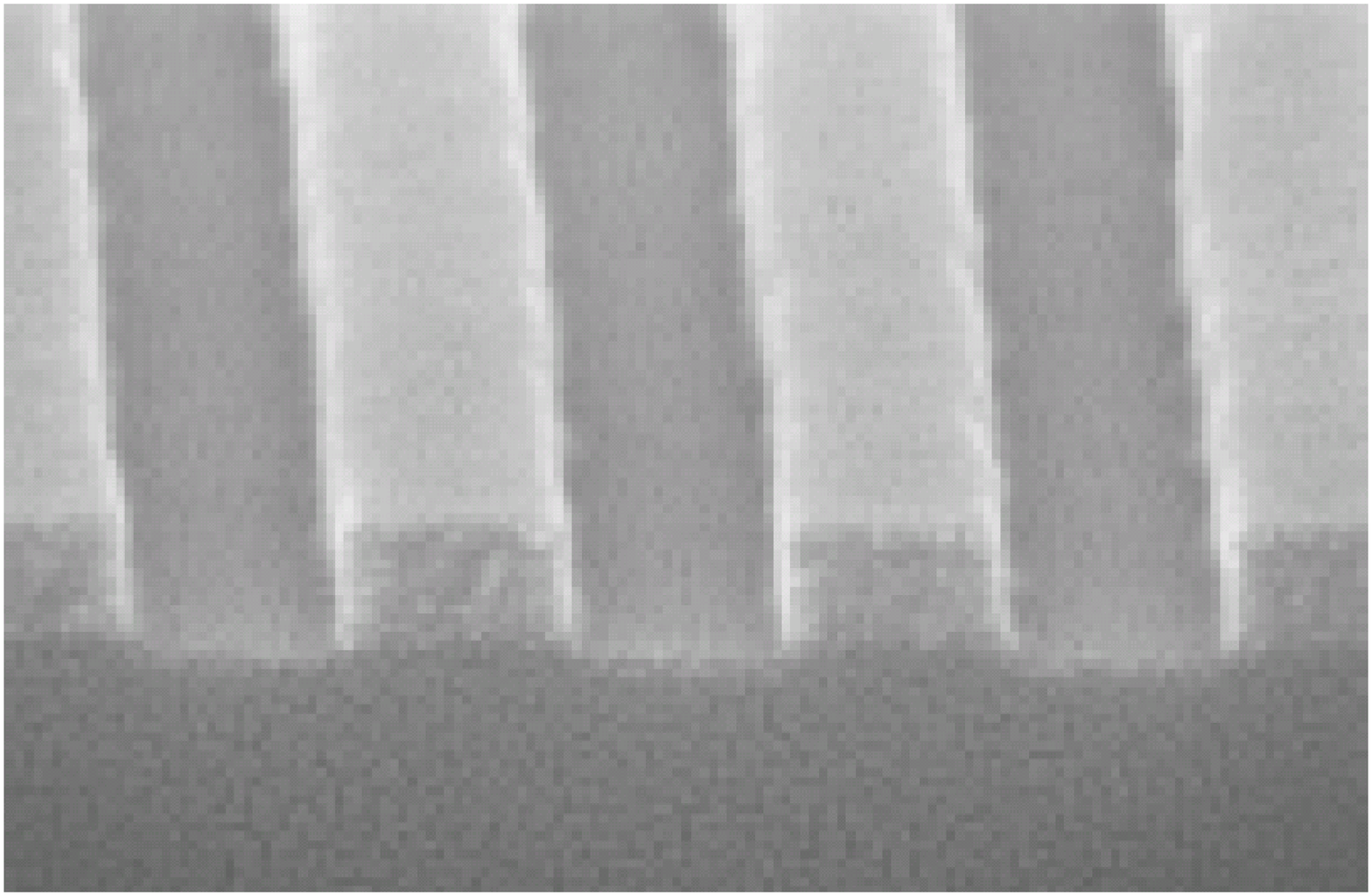}\hfill
\includegraphics[width=4cm,height=2.63cm]{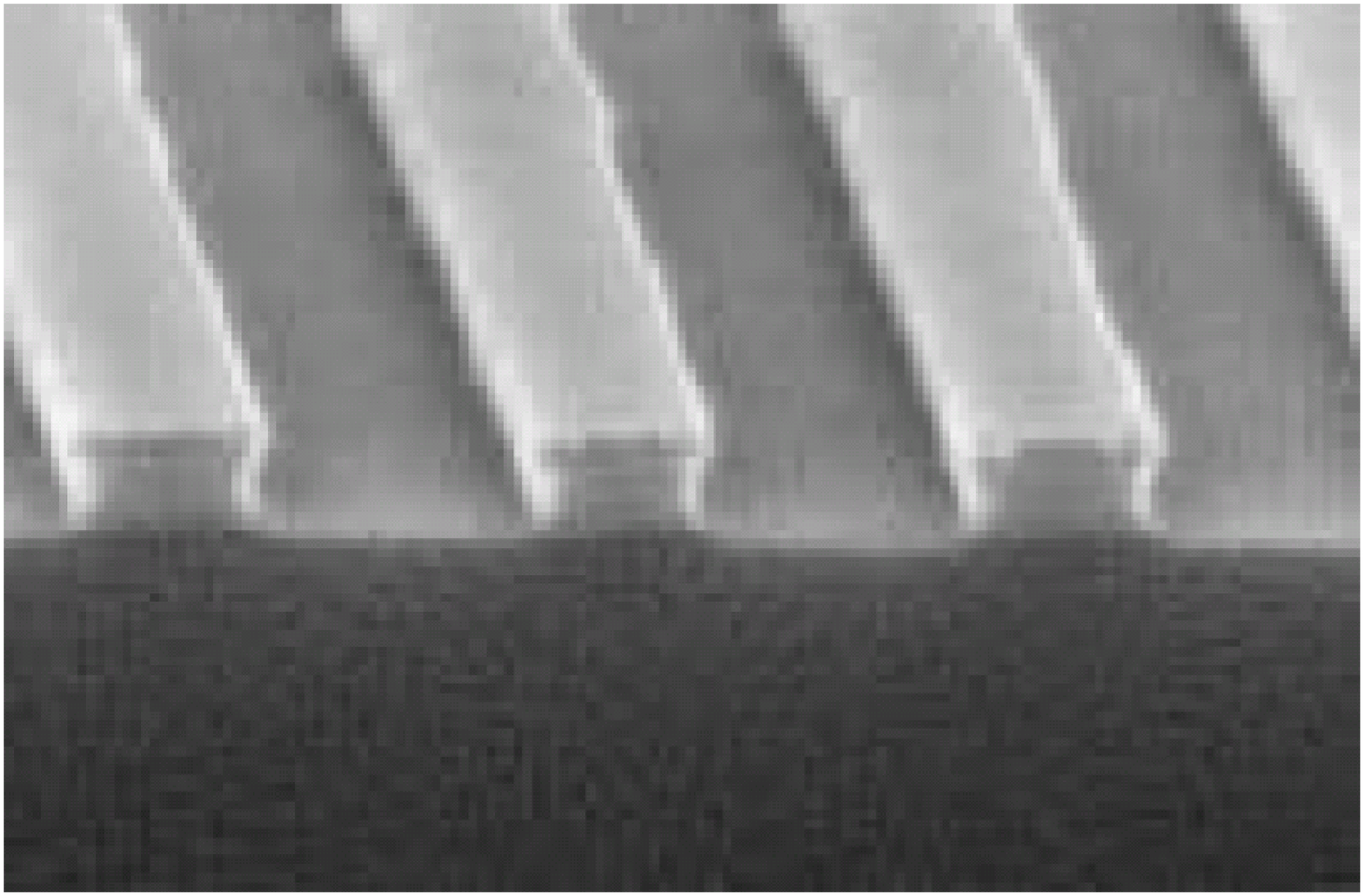}\hfill
\includegraphics[width=4cm,height=2.63cm]{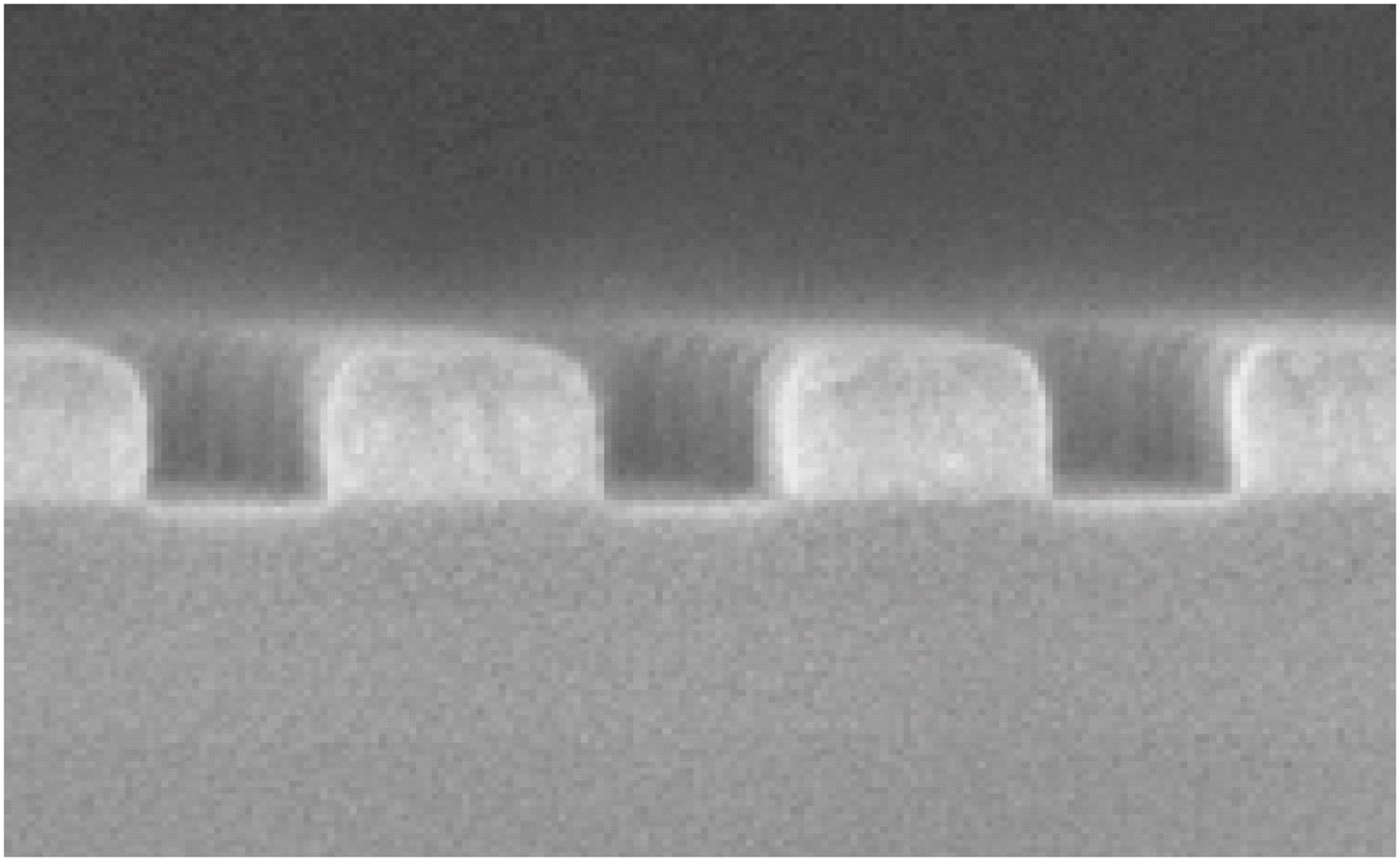}\\
\includegraphics[width=4cm,height=2.63cm]{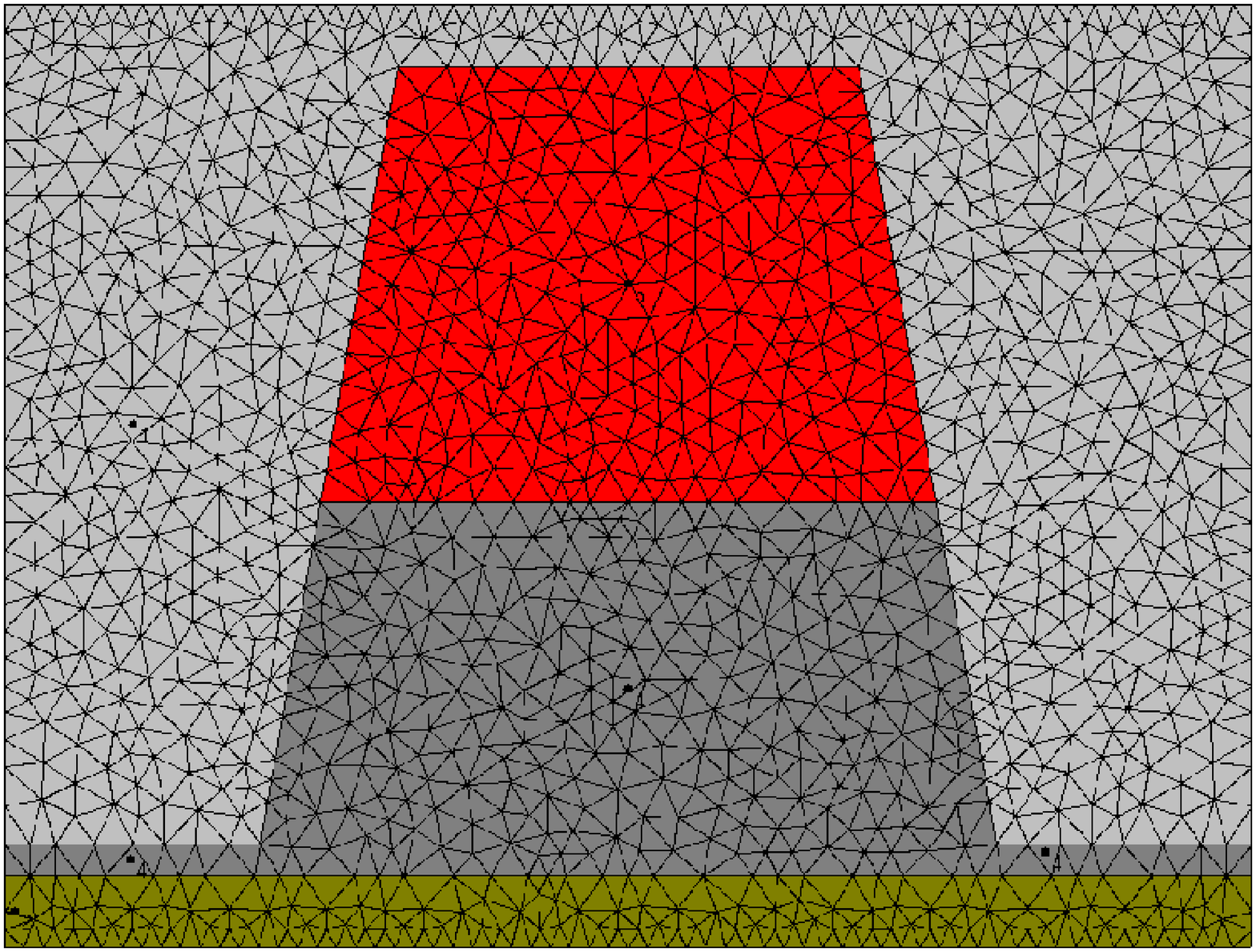}\hfill
\includegraphics[width=4cm,height=2.63cm]{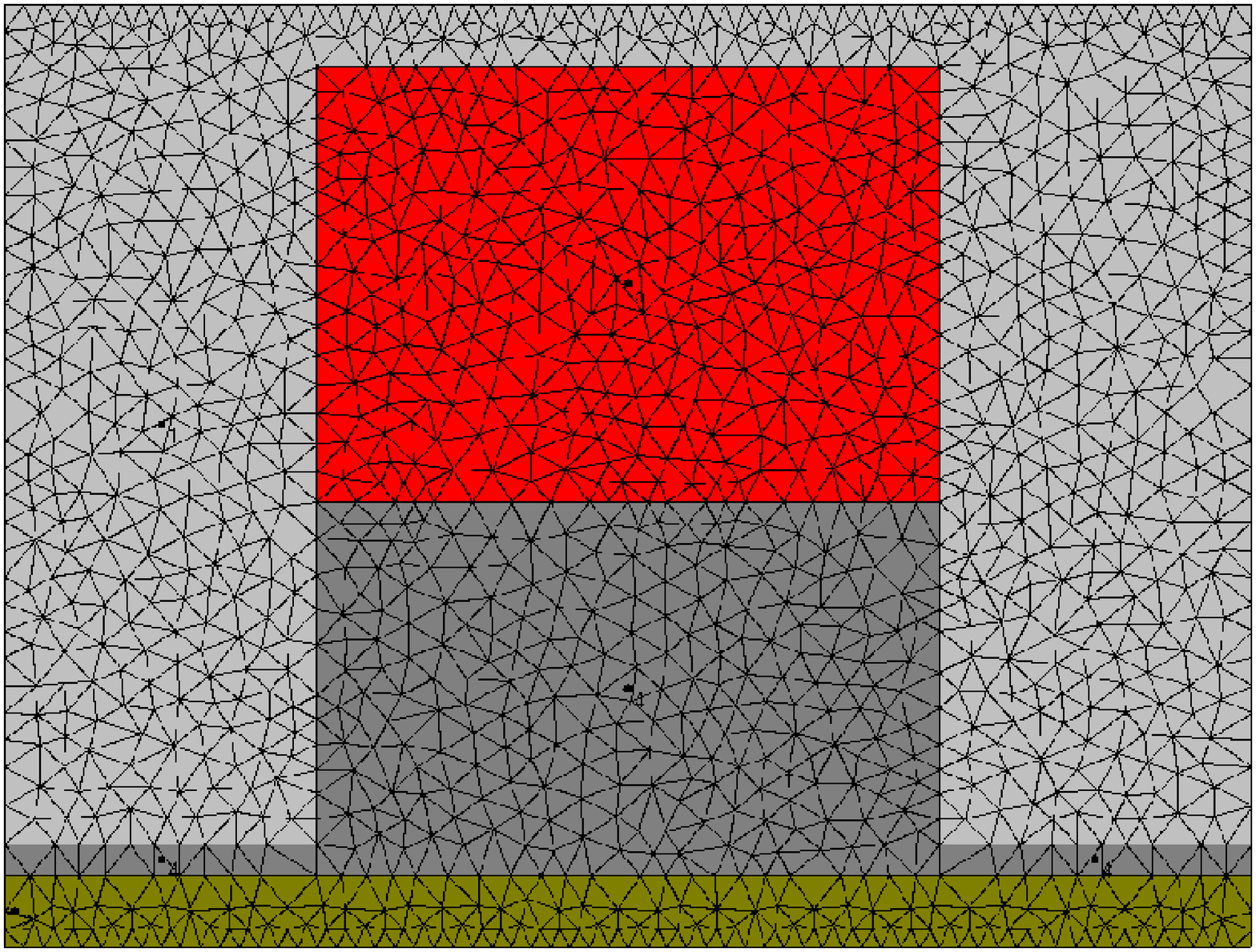}\hfill
\includegraphics[width=4cm,height=2.63cm]{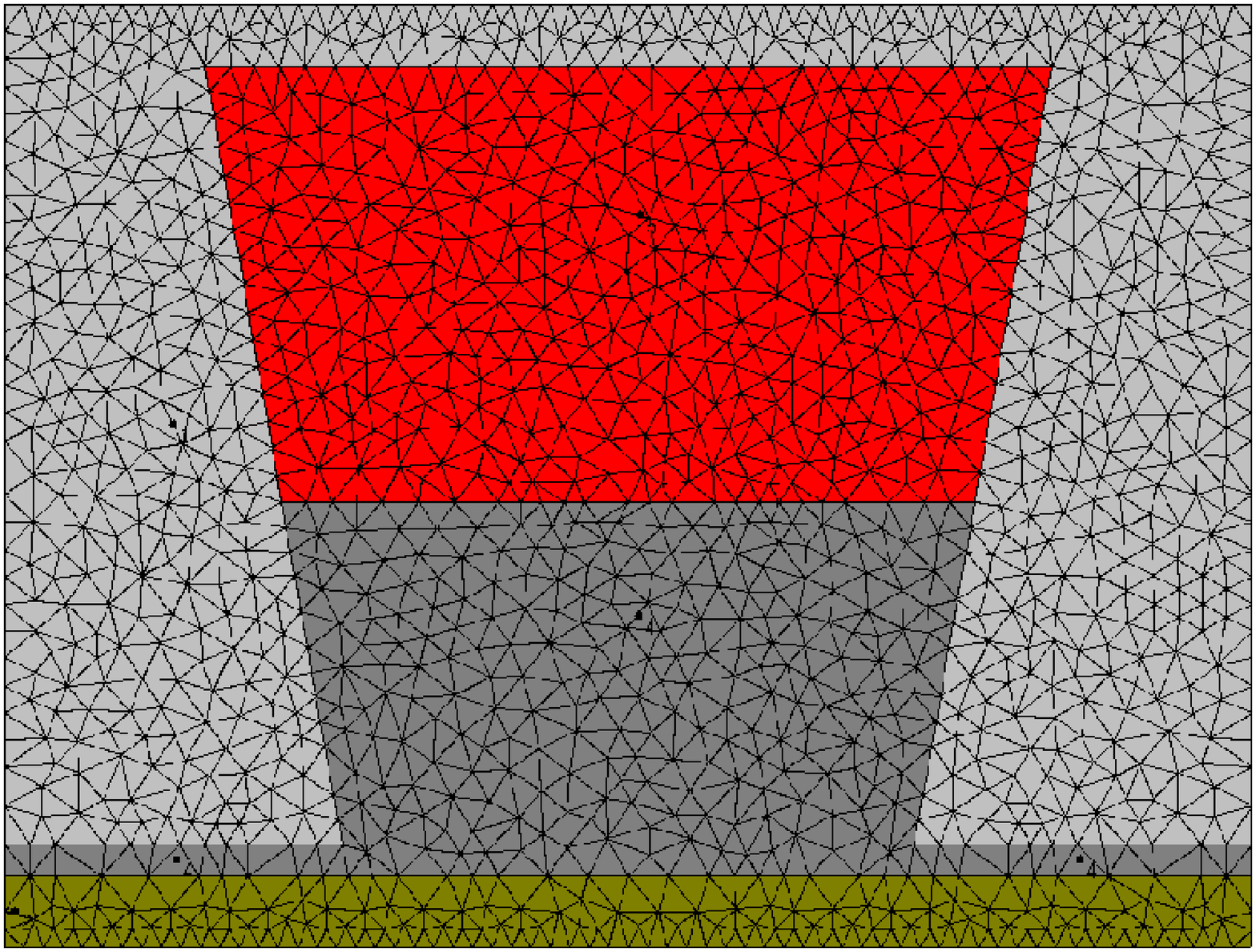}\hfill
\includegraphics[width=4cm,height=2.63cm]{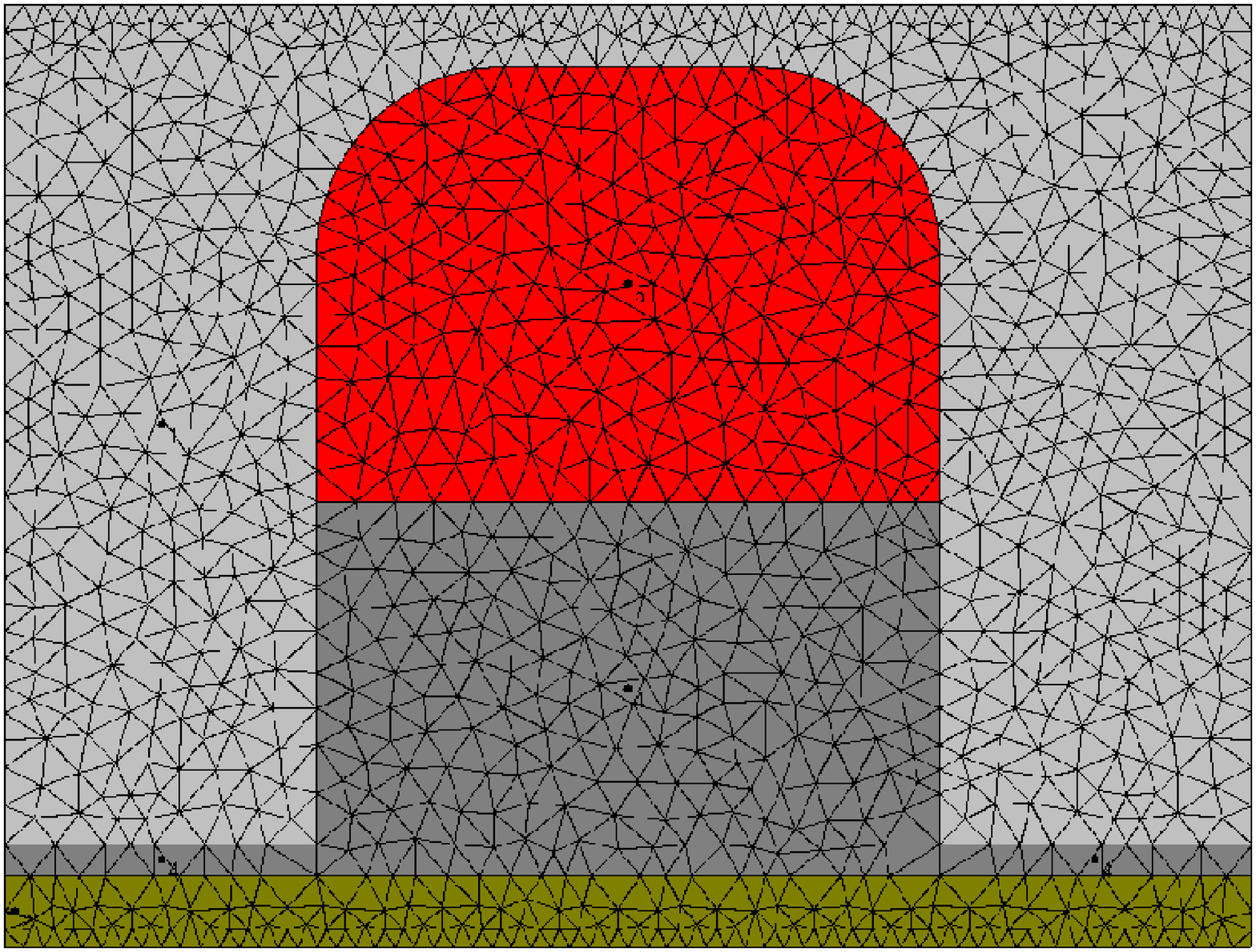}\\
\caption{\label{fig:triang}SEM pictures of EUV mask patterns and corresponding triangulated geometries for FEM computation.}
\end{figure}
Fig. \ref{fig:h4scatter}(b) shows the result of a scatterometry measurement of an EUV test mask (parameters given in Table \ref{table:mask}) considered in the present work.

The positions of the diffraction angles provide information about the pitch of the EUV absorber pattern. However the intensities of the diffraction orders do not carry direct information about other topological features of the mask. The determination of these features from a scatterometry measurement is a challenging inverse problem and a hot topic of actual research. Accurate and fast numerical simulation of the scattering experiment thereby plays a vital role. The experimental situation of single wavelength scatterometry is described by time-harmonic Maxwell's equations for fixed frequency $\omega$. We choose the finite element method to obtain solutions to Maxwell's equations numerically. 
The idea of the finite element method is very simple and will shortly be explained in the following. Maxwell's equations can be transformed without approximations into weak form. Therefore Maxwell's equations for the electric field $\MyField{ E}$ are multiplied by a so called test function ${\MyField{\Phi}}$ and integrated over the computational domain $\Omega$ (e.g. the EUV mask):
\begin{eqnarray}
  \label{eq:MWweak1}
  \int_{\Omega}\left\{{\overline{\MyField{\Phi}}}\cdot\left[\nabla\times{\MyTensor{\mu}}^{-1}\nabla\times\MyField{ E}-\omega^{2}\MyTensor{\epsilon}\, \cdot\MyField{ E}\right]\right\}d^{3}r=0,
\end{eqnarray}
where the bar denotes complex conjugation. With a partial integration (Stoke's theorem) this equation can be transformed to
\begin{eqnarray}
  \label{eq:MWweak2}
&&\int_{\Omega}\left\{\overline{\left(\nabla\times\MyField{\Phi}\right)}\cdot\left({\MyTensor{\mu}}^{-1}\nabla\times\MyField{ E}\right)-\omega^{2}\MyTensor{\epsilon}\, \overline{\MyField{\Phi}}\cdot\MyField{ E}\right\}d^{3}r=\int_{\Gamma}\overline{\MyField{\Phi}}\cdot\MyField{ F} d^{2}r,
\end{eqnarray}
where $\MyField{F}$ contains data of the electric field on the boundary $\Gamma=\partial\Omega$. Defining the complex bilinear form $a$ and linear form $f$:
\begin{eqnarray}
  \label{eq:aDef}
  a(\MyField{\Phi},\MyField{ E})&=&\int_{\Omega}\left\{\overline{\left(\nabla\times\MyField{\Phi}\right)}\cdot\left({\MyTensor{\mu}}^{-1}\nabla\times\MyField{ E}\right)-\omega^{2}\MyTensor{\epsilon}\, \overline{\MyField{\Phi}}\cdot\MyField{ E}\right\}d^{3}r,\\
f(\MyField{\Phi})&=&\int_{\Gamma}\overline{\MyField{\Phi}}\cdot\MyField{ F} d^{2}r,
\end{eqnarray}
Maxwell's equation can be stated compactly in weak formulation (i.e. as a variational problem) \cite{MON03}:
\begin{description}
\item Find $\MyField{ E}\in V=H(curl,\Omega)$ such that:
\begin{eqnarray}
  \label{eq:maxwellWeak}
&&a(\MyField{\Phi},\MyField{ E})=f(\MyField{\Phi})
\,,\;\forall \MyField{\Phi}\in V.
\end{eqnarray}
\end{description} 
$H(curl,\Omega)$ is an infinite dimensional function space and the exact solution to Maxwell's equations is an element of this space. The exact mathematical definition of $H(curl,\Omega)$ and the theory behind the weak formulation can be found e.g. in \cite{MON03}. The finite element method discretizes the infinite dimensional space $V$ by restricting it to a finite dimensional subspace $V_{h}\subset V$. The weak formulation \eqref{eq:maxwellWeak} is then stated on this finite dimensional subspace and therewith discretized. For the performance of the finite element method for Maxwell's equations it is very important to construct $V_{h}$ in such a way that the geometrical properties of $V$ carry over to $V_{h}$. The appropriate construction with so called edge elements e.g. assures the continuity of the solution at boundaries of different materials, a fundamental property of the electric field (in the usual case of absent surface charges). The finite element method does not approximate Maxwell's equations itself ($a(.,.)$ and $f(.)$ are not changed) only the the solution space is approximated with a finite dimensional version $V_{h}$. The construction of $V_{h}$ is done in the following. First the computational domain is subdivided into small  patches, e.g. triangles in 2D or tetrahedrons in 3D. On these patches a local finite elements space is defined which usually consists of polynomial functions (so called shape functions) up to a certain order $p$. Since the electric field is vector valued the shape functions are also vectorial. To increase the accuracy of a finite element solution either the size of the patches can be reduced ($h$ refinement) or the polynomial order can be increased ($p$ refinement). For wave propagation phenomena as described by Maxwell's equations ''high order'' is often a significantly better strategy then ''fine grid'' as we will see in a convergence analysis at the end of this section. 

It has been shown that the FEM method is an excellent choice for this type of application \cite{Burger2005bacus}. It produces very accurate results in relatively short time which is vital for the solution of the inverse problem.
The FEM method has several advantages \cite{Burger2005bacus}:
\begin{itemize}
\item
Maxwell's equations describing the scattering problem are solved rigorously without approximations.
\item 
The flexibility of triangulations allows modeling of virtually arbitrary structures, as illustrated in Fig. \ref{fig:triang}.
\item 
Adaptive mesh-refinement strategies lead to very accurate results and small computational times which are crucial points for application of a numerical method to the characterization of EUV masks.
\item
Choosing appropriate localized ansatz functions for the solution of Maxwell's equations physical properties of the electric field like discontinuities or singularities can be modeled very accurately and don't give rise to numerical problems.
\item
It is mathematically proven that the FEM approach converges with a fixed convergence rate towards the exact solution of Maxwell-type problems for decreasing mesh width of the triangulation. Therefore it is easy to check if numerical results can be trusted.
\end{itemize}
Throughout this paper we use the FEM package JCMsuite developed by the Zuse Institute Berlin and JCMwave for numerical solution of Maxwell's equations. JCMsuite has been successfully applied to a wide range of electromagnetic field computations including waveguide structures \cite{Burger2005a}, DUV phase masks \cite{Burger2005bacus}, and other nano-structured materials \cite{Enkrich2005a,Kalkbrenner2005a}. It provides higher order edge elements, multigrid methods, a-posteriori error control, adaptive mesh refinement, etc. Furthermore a special domain decomposition algorithm implemented in JCMsuite is utilized for simulation of EUV masks \cite{Zschiedrich2005b}. Light propagation in the multilayer stack beneath the absorber pattern can be determined analytically. The domain decomposition algorithm combines the analytical solution of the multilayer stack with the FEM solution of the absorber pattern, decreasing computational time and increasing accuracy of simulation results. The iterative idea of the domain decomposition algorithm is depicted in Fig. \ref{fig:defineParam}(b). In the first step light scattering from the absorber pattern is computed with the finite element method neglecting the multilayer (1). The resulting field on the lower boundary is decomposed into diffraction modes (2) and the reflection of each diffraction mode from the multilayer is computed analytically (3). In the next step the incident field from above and the reflected diffraction modes from below are coupled into the computational domain (i.e. the absorber pattern) and the FEM computation is repeated, yielding a new solution for the electric field. This solution is again decomposed into diffraction modes on the lower boundary and the iteration is continued. This iterative solution converges to the exact solution of the coupled problem \cite{Zschiedrich2005b}, i.e. domain decomposition as implemented in JCMsuite is a ``rigorous'' method \cite{Zsch08}. For the problem class investigated in this paper, few iterations are enough to reach a relative error $<0.1\%$ for the diffraction orders. It is worth noting that the FEM system has to be assembled only once. Also the LU-factorization (i.e. inversion) of the FEM matrix is only performed once and applied to the new right hand side in each iteration step.
\section{Convergence of FEM simulations}
\label{sec:convergence}
\begin{figure}[t]
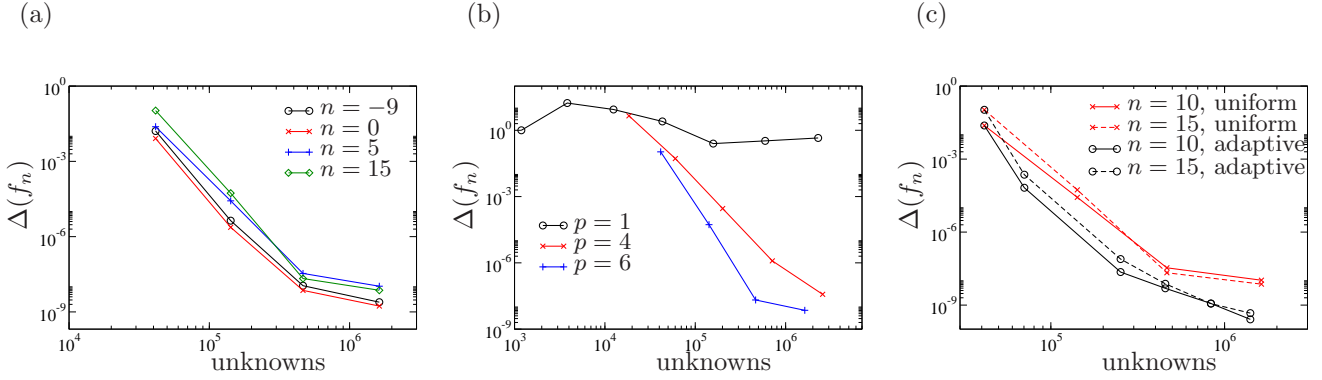

\psfrag{o0}{\small{$n=0$}}
\psfrag{o-9}{\small{$n=-9$}}
\psfrag{o5}{\small{$n=5$}}
\psfrag{o15}{\small{$n=15$}}
\psfrag{fem1}{\small{$p=1$}}
\psfrag{fem4}{\small{$p=4$}}
\psfrag{fem6}{\small{$p=6$}}
\psfrag{unk}{unknowns}
\psfrag{rel}{$\Delta(f_{n})$}
\psfrag{o10no}{\small{$n=10$, uniform}}
\psfrag{o15no}{\small{$n=15$, uniform}}
\psfrag{o10yes}{\small{$n=10$, adaptive}}
\psfrag{o15yes}{\small{$n=15$, adaptive}}
(a)\hspace{5.5cm}(b)\hspace{5.5cm}(c)\vspace{6mm}\\
\includegraphics[width=5.3cm]{fig/convFEM6No.eps}\hfill
\includegraphics[width=5.3cm]{fig/convOrder15FEM.eps}\hfill
\includegraphics[width=5.3cm]{fig/convFEM6AdaNo.eps}\hfill
\caption{{\color{black}\label{fig:conv}Relative error of complex diffraction orders $\Delta(f_{n})$ in dependence of number of unknowns of FEM computation for different diffraction orders $n$, finite element degrees $p$ and refinement strategies. (a) Regular refinement, FEM degree $p=6$; (b) regular refinement, diffraction order $n=15$; (c) finite element degree $p=6$.}}
\end{figure}

 Fig. \ref{fig:conv} shows the convergence of the complex diffraction orders for the simulated EUV mask with Layout H5, see Table \ref{table:mask}. The error is given relative to the most accurately computed solution. Since we look at the complex diffraction orders phase errors are also included. Fig. \ref{fig:conv}(a) shows the relative error of the diffraction orders $n=-9,0,5,15$ in dependence on the number of unknowns for regular refinement and finite element degree $p=6$. All orders are converging at almost the same rate. Fig. \ref{fig:conv}(b) shows the big benefit of higher order elements. For first order finite elements (degree $p=1$) we see almost no convergence, which is in agreement with the slow convergence of first order finite difference time domain (FDTD) methods for photo mask simulations, \cite{Burger2005bacus}. For orders $p=4$ and $p=6$ the convergence rate of the diffraction orders increases dramatically. The flexibility of the FEM method allows to choose optimal settings of grid refinement $h$ and polynomial order $p$ of the ansatz functions to achieve a specified accuracy in minimum computational time. Fig. \ref{fig:conv}(c) shows a comparison of the convergence rate for uniform and adaptive refinement for finite element degree $p=6$ and the diffraction orders $n=10,15$. Adaptive refinement increases the convergence rate even more. For the same number of unknowns the relative error is almost 2 orders of magnitude smaller. In the following computations we used finite element order $p=4$, and a grid refinement level where the relative error of the diffraction orders is below 0.1\%.

\section{Inverse Scatterometric Problem}
\begin{figure}[t]
\psfrag{ein}{$\MyField{E}_{in}$}
\psfrag{xco}{x}
\psfrag{yco}{y}
\psfrag{kin}{$\MyField{ k}_{in}$}
\psfrag{eout}{$\MyField{E_{s}}$}
\psfrag{omega}{$\Omega$}
\psfrag{gamma}{$\Gamma=\partial\Omega$}
\psfrag{r3}{$\mathrm{R}^{2}\setminus\Omega$}
(a)\hspace{8.5cm}(b)\\\vspace{6mm}
\includegraphics[width=6cm]{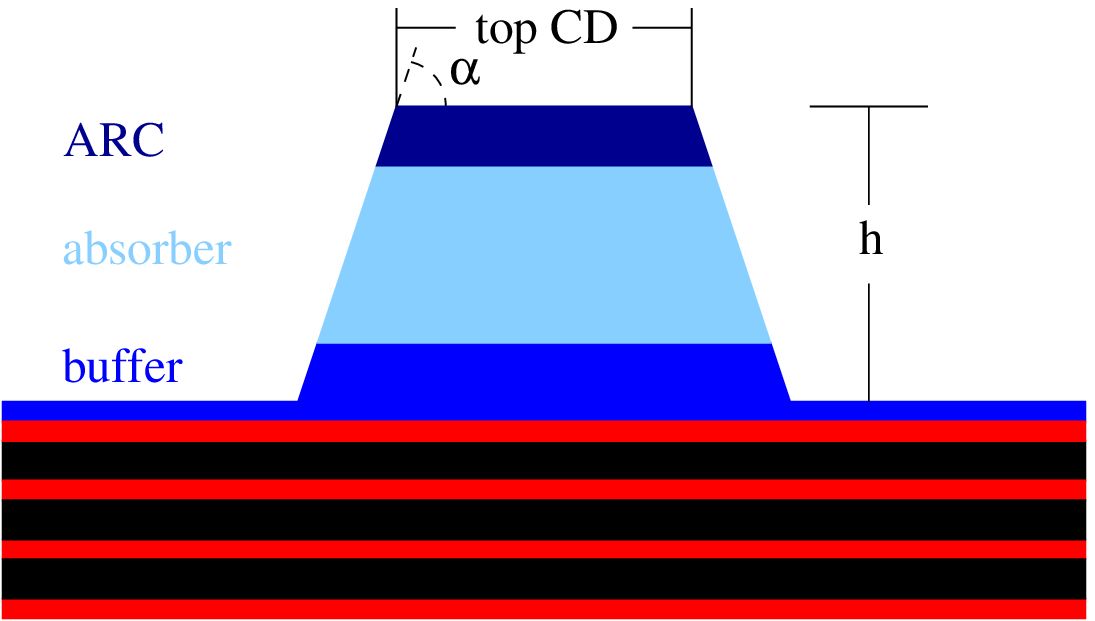}\hfill
\includegraphics[width=6cm]{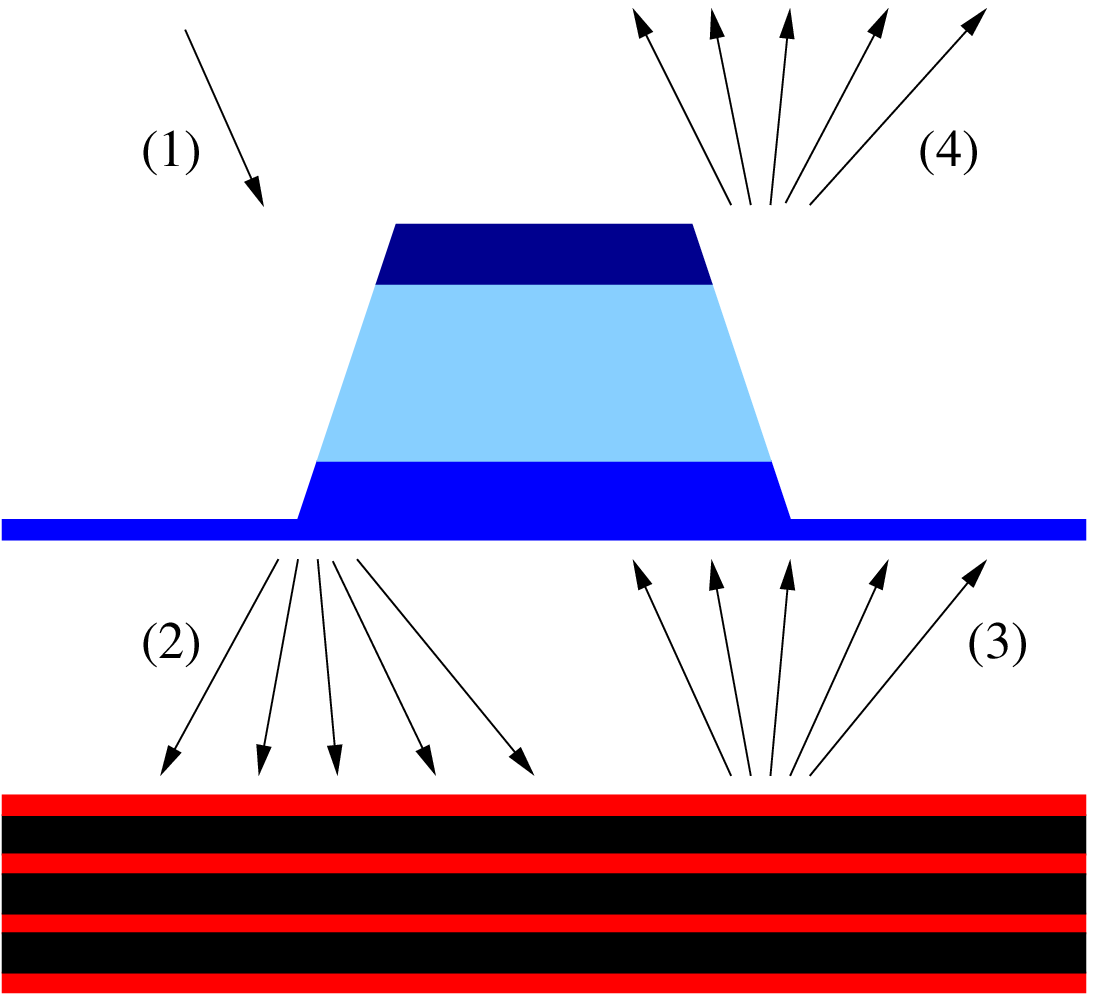}\hfill
\caption{\label{fig:defineParam}(a) Geometrical parameters describing absorber line profile. (b) Illustration of scattering process from EUV mask: (1) incident plane wave, (2) light is decomposed into diffraction orders by absorber line, (3) each diffraction order is reflected with amplitude and phase shift according to multilayer properties, (4) scattered light from EUV mask. Of course also light coming from the multilayer is reflected again by the absorber pattern and also part of the incident light is reflected directly from the absorber pattern.}
\end{figure}
Determining the profile of the periodic absorber lines we have to solve the following optimization problem:
\begin{description}
\item[Optimization problem:] For given experimental diffraction orders $I^{exp}=\left.\left(I^{exp}_{n}\right)\right|_{n=1,\dots,N}$ determine the geometry $G$ of the EUV mask such that:
  \begin{eqnarray}
    \label{eq:optProblem}
G=\arg\; \min\limits_{G'\in\Xi}     d(I^{exp},I^{sim,G'}),
  \end{eqnarray}
where $I^{sim,G'}=\left.\left(I^{sim,G'}_{n}\right)\right|_{n=1,\dots,N}$ are the diffraction orders obtained from the scattering simulation with geometry $G'$. The metric $d$ measures the difference between experimental and simulated diffraction orders. $\Xi$ is the set off all possible geometries.
\end{description}
As stated above the optimization problem is infinite-dimensional and ill-stated. The set of possible geometries $\Xi$ is infinite-dimensional and we only have a finite number $N$ of diffraction orders used for determination of $G$. One can imagine that there are a lot of different geometries which produce similar diffraction patterns. 
\pagebreak

The following questions and problems arise solving \eqref{eq:optProblem}:
\begin{itemize}
\item Experimental diffraction orders $I^{exp}$ are measured with uncertainty.
\item The material properties (permittivity) are not known exactly.
\item The layout of the multilayer beneath the absorber line is not known exactly.
\item How do we restrict $\Xi$ to arrive at a better conditioned problem?
\item Which error measure do we use for $d\,$?
\end{itemize}
Having these points in mind we can not expect that we find a geometry $G_{real}$ such that $d(I^{exp},I^{sim,G_{real}})=0$ and probably a lot of geometries produce errors $d$ of similar magnitude. Consequently the question of accuracy and error bounds for the minimizing geometry $G_{opt}$ of \eqref{eq:optProblem} arises.

The experimental diffraction orders are measured with a relative accuracy of $1\%$. Noise in the EUV detector is of the order of $5\cdot 10^{-5}$ relative to the incident light. Experimental diffraction orders with intensity below this limit are therefore not used for optimization since their relative error is above $100\%$. For $d$ we use the sum of the squares of the relative errors over all diffraction orders. In the present work experimental orders were taken at three different wavelengths for each absorber pattern. All of the orders (with intensity $>0.5\cdot 10^{-5}$) are included computing $d$:
\begin{eqnarray}
  \label{eq:errorDef}
  d(I^{exp},I^{sim,G})=\sum\limits_{n=1}^{N}\left(\frac{I_{n}^{exp}-I_{n}^{sim,G} }{I_{n}^{exp}}\right)^{2}
\end{eqnarray}
For definition of the set of geometries $\Xi$ considered in minimizing \eqref{eq:optProblem} we parameterize the absorber line as depicted in Fig. \ref{fig:defineParam}(a). We have three discrete parameters: the height of the absorber stack $h$, the absorber sidewall angle $\alpha$ and the width of the absorber line, e.g. given at the top, i.e. top CD. Here we will only look at the results for the CD and $\alpha$. The pitch $p$ of the periodic lines can be deduced from the position of the diffraction peaks. The subset of possible geometries $\Xi_{h}\subset\Xi$ is therewith defined.

Accurate modeling of the multilayer is a crucial part in the reconstruction of the absorber pattern \cite{Pomplun2007}. We think that in our setting it has the greatest influence on reconstruction results and is the main cause for uncertainty of the reconstructed parameters. Other error sources are e.g. numerical errors of the FEM simulations. They can be virtually neglected using appropriate computational settings after a convergence analysis, according to Section \ref{sec:convergence}. In \cite{Gross08} the uncertainty of the reconstruction results caused by measurement errors of the diffraction orders is analyzed and quantified, which is also important to judge the reliability of reconstructed geometry.

The experimental orders of our analysis have a measurement uncertainty of $1\%$. The diffraction orders of the numerically determined best fitting geometries however still differ much more from the experimental orders then these measurement errors, see Fig \ref{fig:ordersVGL}. Consequently at this stage we focus on the reconstruction error caused by inaccurate modeling of the multilayer.

\section{Modeling of the Multilayer}
\begin{figure}[t]
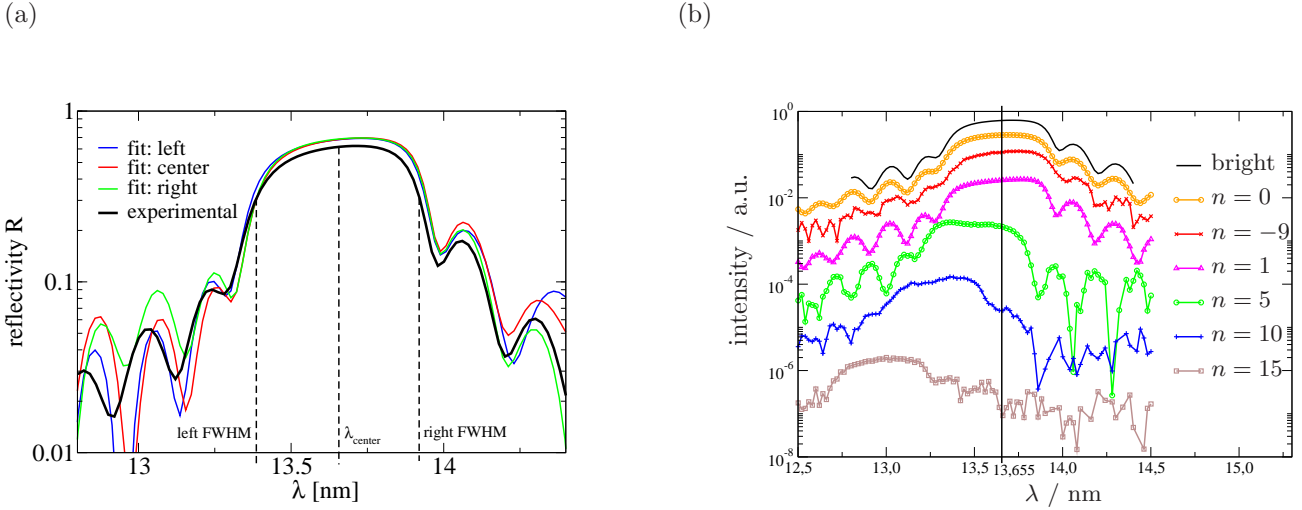

\psfrag{hell}{\small{bright}}
\psfrag{o0}{\small{$n=0$}}
\psfrag{o-9}{\small{$n=-9$}}
\psfrag{oE}{\small{$n=1$}}
\psfrag{o5}{\small{$n=5$}}
\psfrag{o10}{\small{$n=10$}}
\psfrag{o15}{\small{$n=15$}}
\psfrag{lambda}{\small{$\lambda  $ / nm}}
\psfrag{int}{intensity / a.u.}
\psfrag{num}{numerical fit}
\psfrag{exp}{experimental bright field}
\psfrag{refl}{Reflectivity}
(a)\hspace{8.5cm}(b)\\\vspace{6mm}\\
\includegraphics[width=7.5cm]{fig/MLCurvesH4.eps}\hfill
\includegraphics[width=7.5cm]{fig/ordersLambdaH4.eps}\hfill
\caption{{\color{black}\label{fig:ordersLambda}(a) Different numerical fits of experimental bright field curve of field H4. For each curve a different wavelength interval of the measured bright field curve was fitted: left to center wavelength, around center wavelength and right to center wavelength, (b)  Bright field measurement and diffraction orders $n$ of field H4 in dependence on wavelength. Intensity of diffraction orders are scaled for better comparison. The wavelength $\lambda=13.655\,$nm corresponds to the center wavelength of the bright field curve.}}
\end{figure}
\label{sec:modellingMultilayer}
The scatterometric measurements of the EUV mask were performed close to the later operating wavelength of the EUV mask. Fig. \ref{fig:ordersLambda}(a) shows a typical reflectivity curve of a blank multilayer stack. A maximum reflectivity of 60\% to 70\% is reached when the component of the $k$-vector of the incident plane wave which is orthogonal to the layers is in resonance with the half period of the multilayer. These so called bright field curves were measured on unstructured regions close to each of the test fields of the EUV mask. Each scatterometric measurement was performed for the left and right full-width half-maximum (FWHM)-wavelength and center wavelength $\lambda_{center}$ (mean of right and left FWHM wavelength), see Fig. \ref{fig:ordersLambda}(a).

The usage of resonant EUV light has a great advantage. The intensity of reflected experimental diffraction modes is increased significantly since the EUV stack reflects these modes well in the resonant case. This makes it possible to measure more diffraction orders above the limiting intensity of the EUV detector. Especially the high diffraction orders carry the most information about the absorber pattern \cite{Pomplun2007} and are therefore important for a well conditioned reconstruction. 

Light scattering from an EUV mask can be decomposed according to Fig. \ref{fig:defineParam}(b). It follows that the intensities of the diffraction modes used for reconstruction depend on the shape of the absorber pattern and the multilayer, which is demonstrated in Fig. \ref{fig:ordersLambda}(b). The wavelength dependency of the diffraction orders resemble the bright field curve to a great extend. The disadvantage of using resonant EUV light is obvious. If the multilayer is not modeled correctly the absorber pattern can not be reconstructed accurately. Therefore experimental data (the bright field curve) is needed and has to be fitted by modeling the multilayer before reconstruction.

Since the multilayer curves for all test fields had a similar shape only bright field curves from two different fields (H4 and H5) were used modeling the multilayer. For both of these curves the multilayer profile was computed by fitting simulated reflectivity to the experimental values. As material stack a Molybdenum/Silicon (MoSi) multilayer with MoSi2 interlayers was used. Diffusive or rough interfaces between the materials were not taken into account. Optimizing a complete measured reflectivity curve over a large wavelength interval was a difficult task and the results were not very satisfactory for the EUV test mask. Difficulties in the modeling are the mentioned interlayers and the unknown thickness profile of the multilayer. When growing a multilayer it is hard to determine how well the nominal values of the multilayer are reached. Thinking of a multilayer with e.g. 50 bilayers with inter- or diffusion layers of varying thickness one arrives at a optimization problem with several hundreds of unknowns. Consequently modeling of the multilayer itself is a difficult task and actual research problem \cite{Scholze06}.

Here we decided to model three small wavelength intervals of the complete multilayer bright field curve independently, left to the center wavelength, around the center wavelength, right to center wavelength. In total this gives 6 different multilayer curves (3 for H4 and H5 respectively) which were used for reconstruction of the absorber profile. The fits of the bright field curve for field H4 are shown in Fig. \ref{fig:ordersLambda}(a). The ``fit: right'' curve fits the right side wiggles of the experimental curve best, the ``fit: left'' the left side wiggles. Although fitting only a small wavelength interval we still see rather poor agreement for the side wiggles for all fits. 

In total we get 6 reconstructed geometries, one for each modeled multilayer. These are compared to quantify the uncertainties of the reconstruction process. Although the experimental bright field curves of the considered test fields had a similar shape their center wavelengths differed slightly, see Table \ref{table:lambdaCenter}. Scaling all layers of a modeled multilayer equally, each of the 6 bright field curves used for reconstruction could be shifted to match the center wavelength of the field it is used for. We also compare reconstruction results with and without this shift.
\begin{table}[h]
\begin{centering}
\begin{tabular}{lcc}
Test field& center wavelength [nm]\\
\hline
H1 & 13.582\\
H2 & 13.615\\
H3 & 13.639\\
H4 & 13.655\\
H5 & 13.664\\
\end{tabular}\hspace{3cm}
\begin{tabular}{lcc}
Test field& center wavelength [nm]\\
\hline
D4 & 13.6479\\
H4 & 13.6510\\
F6 & 13.6769\\
D8 & 13.6470\\
H8 & 13.6518\\
\end{tabular}
\end{centering}
\caption{\label{table:lambdaCenter}Center wavelength of bright field curves of test fields.}
\end{table}
\section{Reconstruction Results}
\label{sec:results}
\begin{figure}[t]
\centering
\includegraphics[width=14cm]{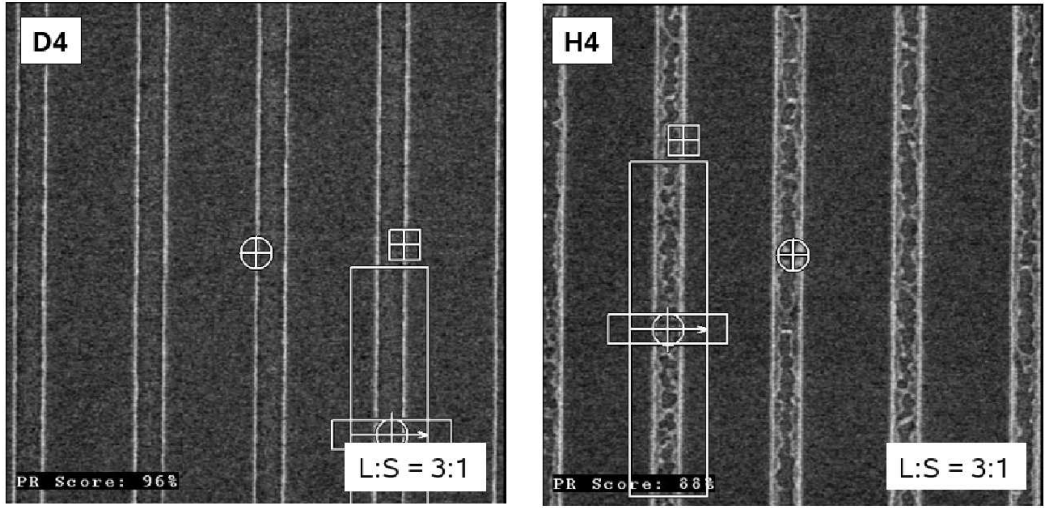}
\caption{{\color{black}\label{fig:etching}Top down SEM image of uniformity fields D4 and H4. It can be seen that the lines for field H4 are not completely etched.}}
\end{figure}

We applied the described reconstruction procedure to the test fields with nominal values given in Section \ref{sec::scatterometry}. The profiles of the absorber lines were measured with direct AFM and CD-SEM for a quantitative evaluation of the indirect reconstruction results. The microscopically measured values are given in table \ref{table:profileMicros} and \ref{table:profileMicrosAFM}. We notice that the CD of the uniformity field of H4 differs significantly from that of the other fields although the nominal value of $180\,$nm is the same for all fields. Fig. \ref{fig:etching} shows the explanation, the lines for the uniformity field H4 are not etched to the bottom. We can already expect poor reconstruction results for this field \cite{Scholze07}.
\begin{table}[h]
\begin{centering}
\begin{tabular}{lcc}
test field & CD [nm]& CD uniform. [nm] \\
\hline
 H1  &   138.3 &   2.5\\
 H2  &   134.3 &   4.0\\
 H3  &   127.8 &   3.8\\
 H4  &   129.8 &   7.1\\
 H5  &   130.4 &   5.5\\
\end{tabular}\hfill
\begin{tabular}{lcc}
test field & CD [nm] & CD uniform. [nm] \\
\hline
 D4 &    186.7  &  3.2\\
 H4 &    168.0  &  7.3\\
 F6 &    180.3  &  2.7\\
 D8 &    188.7  &  3.0\\
 H8 &    184.1  &  3.6\\
\end{tabular}
\end{centering}
\caption{\label{table:profileMicros}SEM measurements of line profiles. CD uniformity is defined as $3\sigma$ over an ensemble of measurements, the CD is the mean over the ensemble}
\end{table}
\begin{table}[h]
\begin{centering}
\centering
\begin{tabular}{lcccc}
test field & CD [nm] (thr. A)& CD [nm] (thr. B)&$\alpha$[\grad] (rising)  & $\alpha_{2}$[\grad] (falling)\\
\hline
 H1  &   169.0 &   160.8   &	78.7&		83.8\\
 H2  &   160.9 &   157.5   &	87.2&		86.5\\
 H3  &   154.7 &   152.2   &	87.2&		86.9\\
 H4  &   155.9 &   153.7   &	87.6&		86.9\\
 H5  &   155.5 &   153.8   &	87.5&		85.9\\
\end{tabular}
\end{centering}
\caption{\label{table:profileMicrosAFM}AFM measurements of line profiles. Sidewall angles were measured in rising and falling mode. The CD was measured at two different AFM thresholds. For uniformity fields no AFM measurements were performed due to the small bright line width.}
\end{table}

\begin{figure}[ht]
(a)\hspace{8.5cm}(b\vspace{9mm})\\
\includegraphics[width=7cm]{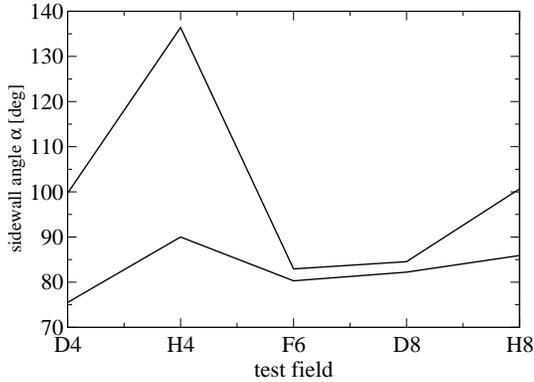}\hfill
\includegraphics[width=7cm]{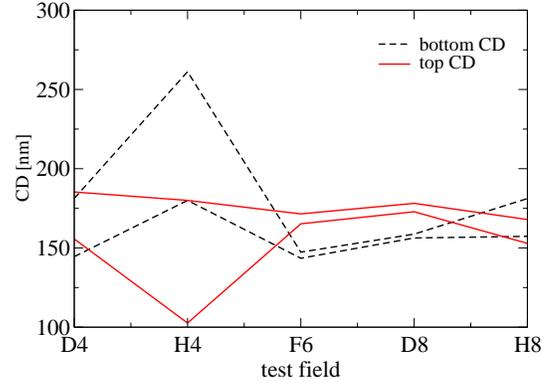}\hfill
\caption{\label{fig:minmax}Minimum and maximum (a)  sidewall angles and (b) top and bottom CDs obtained from reconstruction with 6 different multilayer curves.}
\end{figure}

\begin{figure}[ht]
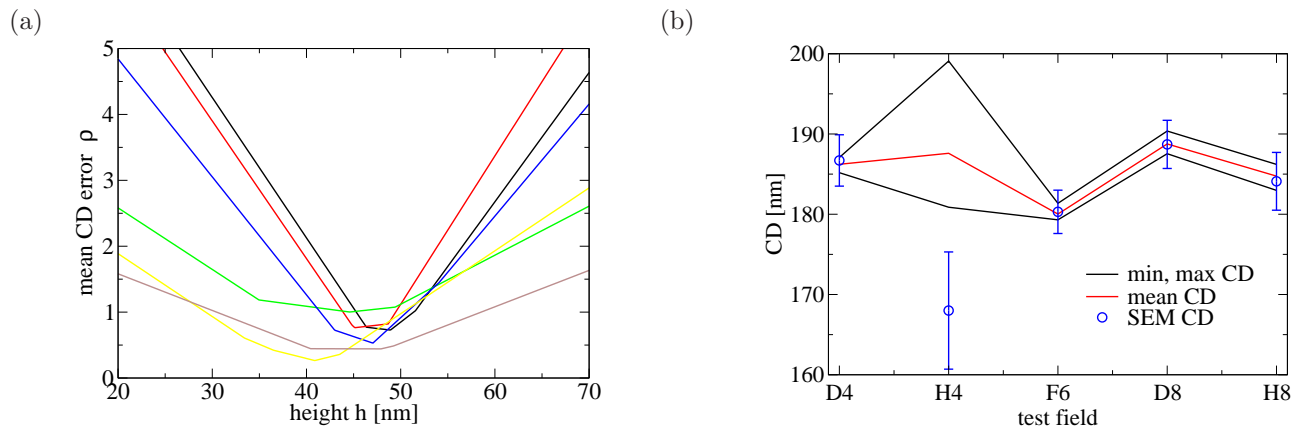

(a)\hspace{8.2cm}(b)\\
\phantom{.........}\includegraphics[width=7cm]{fig/rhoCorr.eps}
\phantom{.....................}\includegraphics[width=7cm]{fig/minmaxCDSEMVgl.eps}
\caption{\label{fig:minmaxCDSEMVgl}(a) Mean error $\rho$ \eqref{eq:correlation} of CD (over all uniformity test fields) between inverse reconstruction and SEM measurements in dependence on absorber height $h$, from which the reconstructed CD was taken. (b) Minimum maximum and mean CD at $\langle h_{opt}\rangle$ from indirect reconstruction with 6 different multilayers, CD SEM values with CD uniformity (with error bars).}
\end{figure}

First we look at the reconstruction results of the uniformity fields (D4, H4, F6, D8, H8) with $180\,$nm bright and $540\,$nm dark lines. Fig. \ref{fig:minmax}(a) shows the range of sidewall angles we obtain when using the 6 different multilayer curves with scaling to match $\lambda_{center}$ of each test field (see Section \ref{sec:modellingMultilayer}). The sidewall angles differ over a large range. Since the bright lines of the uniformity fields were to small for AFM measurements we have only experimental results for the absorber angle of fields H1 to H5, Table \ref{table:profileMicrosAFM}. However it is reasonable to assume that the sidewall angle of the uniformity fields is similar. With this assumption the reconstructed sidewall angles of the uniformity fields differ significantly from the experimental values. Fig. \ref{fig:minmax}(b) shows that also the top and bottom CD differ significantly from the CD SEM values and except from test field F6 and D8 using 6 different multilayers produces CDs over a large range. Field H4 shows the greatest variations as expected because the bright lines were not etched through to the bottom, Fig. \ref{fig:etching}.
 At first glance the results look very disappointing. With a sidewall angle $\not = 90\grad$ the CD of an absorber line however depends on the height in which we measure it $CD=CD(h)$. The question is if we can find a height $h_{opt}$ where the reconstructed CDs are in best agreement with the microscopically determined SEM CDs. Fig. \ref{fig:minmaxCDSEMVgl}(a) depicts the mean CD error $\rho$ (the mean was taken over the 5 test fields) between reconstructed and SEM CDs in dependence on the height $h$ for the 6 modeled multilayers. The mean error $\rho$ is defined as:
\begin{eqnarray}
  \label{eq:correlation}
\rho(h)=\frac{1}{\mbox{\#test fields}}  \sum\limits_{\mbox{test field $i$}}\left|\left(CD_{i}^{SEM}-\langle CD^{SEM}\rangle\right)-\left(CD_{i}^{Inv}(h)-\langle CD^{Inv}(h)\rangle\right)\right|,
\end{eqnarray}
where $CD_{i}^{SEM}$ is the CD determined by SEM and $CD_{i}^{Inv}(h)$ the CD at height $h$ of the absorber line determined by inverse reconstruction. The mean value $\langle \dots\rangle$ is taken over all test fields of the summation, i.e. $\{H1,H2,H3,H4,H5\}$ and $\{D4,H4,F6,D8,H8\}$ respectively. In the definition of $\rho$ a constant offset between both methods is neglected. Also the CD measured by AFM and SEM differ by an offset. Looking at Fig. \ref{fig:minmaxCDSEMVgl}(a) we see that although producing wrong top and bottom CDs the reconstructed CDs at the height of $40\,$nm to $50\,$nm agree very well with the SEM CDs for all used multilayers. This is slightly above the half of the total absorber height. For each multilayer $h_{opt}$ (which minimizes $\rho$) was determined. It would be desirable to analyze more test fields, e.g. with absorber lines of different heights to check if $h_{opt}$ is always at the same position $\bar h_{opt}$ relative to the total height of the line. If that would be the case one could use this value $\bar h_{opt}$ for all reconstructions and would not need prior CD SEM measurements to minimize $\rho$ and therewith determine $h_{opt}$. We take the mean of the six obtained $h_{opt}$ values: $\bar h_{opt}=\langle h_{opt}\rangle$. Of course an analysis of a greater number of test fields could give a better estimation for $\bar h_{opt}$. For a quantification of the uncertainty of the reconstructed CDs we now determine the CDs at height $\bar h_{opt}=\langle h_{opt}\rangle$ from our simulations with 6 different multilayers. To clarify again: for each of the 6 modeled multilayers we get a different $h_{opt}$, which can only be determined using the microscopical results for the CD. Since in practice we want to use the indicrect method without SEM measurements we have to choose $\bar h_{opt}$. Our experience has shown to choose it slighlty above the half absorber height. With fixed $\bar h_{opt}$ the CDs from the simulations with the different multilayers will not be in best agreement with the SEM results and we will utilize this disagreement for our error estimations.
The minimum, maximum and mean CDs of the 6 simulations with different multilayers at same fixed $\bar h_{opt}$ are shown in Fig. \ref{fig:minmaxCDSEMVgl}. The reconstructed results and error bounds (minimum and maximum CD at $\bar h_{opt}$) are well within the error bounds of the CD SEM results which demonstrates the robustness of the reconstruction of the CD very nicely. Field H4 again shows larger disagreement. Surprisingly the values obtained from the indirect reconstruction even for this field seem more reasonable than the CD SEM results, Fig. \ref{fig:etching} shows that the lines are not etched through but still have same CD.

\begin{figure}[ht]
(a)\hspace{8.3cm}(b)\vspace{5mm}\\
\phantom{.........}\includegraphics[width=7cm]{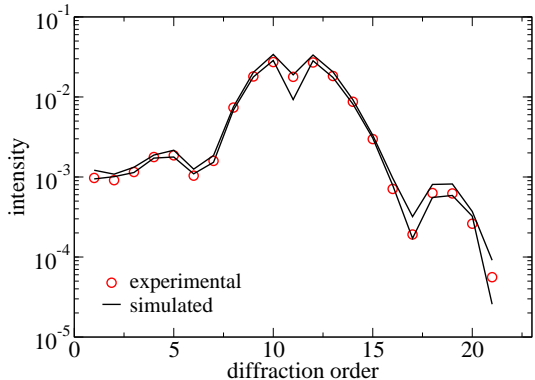}\hfill
\phantom{.....................}\includegraphics[width=7cm]{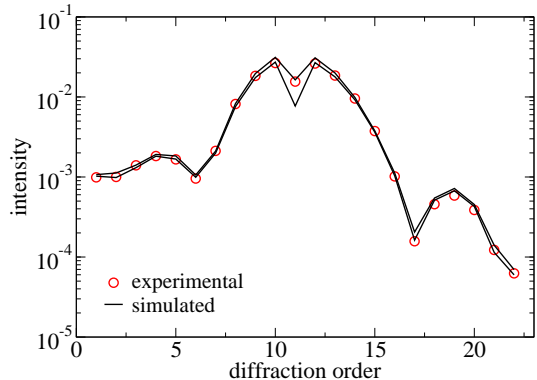}
\caption{\label{fig:ordersVGL}Experimental and simulated diffraction orders of reconstructed geometry. The minimum and maximum diffraction order over all 6 modeled multilayers is given; (a) field H4 with incident wavelength = left FWHM wavelength, (b) field F6 with incident wavelength = center wavelength.}
\end{figure}

To summarize the sidewall angles of the absorber lines can not be determined very well due to inaccurate modeling of the multilayer, however the CD at the middle of the absorber line can be reconstructed very well. 
Fig. \ref{fig:ordersVGL} shows the diffraction orders which were computed for the best fitting geometries for two example test fields. Since we used 6 different multilayers we also get 6 intensities for each diffraction mode. Fig. \ref{fig:ordersVGL} shows the minimum and the maximum of these diffraction intensities. We notice that almost all experimental diffraction orders are located between the smallest and largest simulated diffraction orders. A better modeling of the multilayer is therefore the next step for improving the reconstruction performance, e.g. for the sidewall angle. The measurement uncertainties are still much smaller than the tube spanned by the simulated orders, and therewith it is confirmed that the measurement uncertainties are of minor importance.

For the $140\,$nm lines we compared the results obtained with and without scaling of the multilayer with respect to the central wavelength. The results for the CD at $h_{opt}$ are shown in Table \ref{table:compareHFields}. Here we performed the reconstruction with only one fixed multilayer model. With scaling the relative errors are below 0.5\%. The absolute error is below $0.7\,$nm. To our knowledge this is the most accurate CD reconstruction for periodic EUV line masks. Without scaling the relative error of the reconstructed CDs is large and $>1\%$ for some fields.
\begin{table}[h]
\begin{centering}
\begin{tabular}{lccc}
test field & CD$^{SEM}$ & CD$^{Scatt.}$ ($\Delta_{CD}$), with scaling& CD$^{Scatt.}$ ($\Delta_{CD}$), without scaling\\
\hline
 H1  &   138.3 & 136.5 (1.3\%) & 138.9 (0.4\%)\\
 H2  &   134.3 & 133.1 (0.9\%) & 134.0 (0.2\%)\\
 H3  &   127.8 & 129.4 (1.3\%) & 127.9 (0.1\%)\\
 H4  &   129.8 & 131.1 (1.0\%) & 129.3 (0.4\%)\\
 H5  &   130.4 & 130.4 (0.0\%) & 130.4 (0.0\%)\\
\end{tabular}
\end{centering}
\caption{\label{table:compareHFields}Comparison of CD-SEM measurements and reconstructed CDs (given in nm) with scaled and unscaled multilayer at height $h_{opt}$ (height with maximal correlation of CD to SEM CD). The relative error $\Delta_{CD}$ is given in brackets.}
\end{table}

\section{Conclusions}
We have shown that EUV scatterometry in combination with FEM simulations is a very well suited method for profile reconstruction of EUV masks. A comparison of CD values obtained by inverse reconstruction and direct microscopical SEM measurements shows an agreement below $1\,$nm. The diffraction intensities of a EUV mask depend on the actual absorber structure and the properties of the underlying multilayer. Without an accurate modeling of the multilayer not all parameters of the absorber profile can be reconstructed, e.g. the sidewall angle. However the performance of the reconstruction of the CD is very good and robust. By modifying the multilayer uncertainty estimations in the reconstruction results of the CD can be obtained. Those are comparable to uncertainties of the CD SEM measurements.

Differences in the diffraction intensities between the best fitting geometries and the experimental values can also be explained by the difficulties of modeling the multilayer with high accuracy. This will be the next step in improving reconstruction results.

\label{sec::conclusions}
\bibliography{/home/numerik/bzfpompl/myBib}

\begin{thebibliography}{10}

\bibitem{Sugawara05a}
M.~Sugawara, I.~Nishiyama, and M.~Takai, ``Influence of asymmetry of diffracted
  light on printability in {EUV} lithography,''  {\bf 5751}, pp.~721--732,
  Proc. SPIE, 2005.

\bibitem{Sugawara05b}
M.~Sugawara and I.~Nishiyama, ``Impact of slanted absorber sidewall on
  printability in {EUV} lithography,''  {\bf 5992}, Proc. SPIE, 2005.

\bibitem{Ulm98}
G.~Ulm, B.~Beckhoff, R.~Klein, M.~Krumrey, H.~Rabus, and R.~Thornagel, ``The
  {PTB} radiometry laboratory at the {BESSY} {II} electron storage ring,''
  {\bf 3444}, pp.~610--621, Proc. SPIE, 1998.

\bibitem{Scholze05}
F.~Scholze, C.~Laubis, C.~Buchholz, A.~Fischer, S.~Pl{\"o}ger, F.~Scholz,
  H.~Wagner, and G.~Ulm, ``Status of {EUV} reflectometry at {PTB},''  {\bf
  5751}, pp.~749--758, Proc. SPIE, 2005.

\bibitem{Scholze03}
F.~Scholze, J.~T{\"u}mmler, and G.~Ulm, ``High-accuracy radiometry in the {EUV}
  range at the {PTB} soft {X}-ray radiometry beamline,''  {\bf 40},
  pp.~224--228, Metrologia, 2003.

\bibitem{MON03}
P.~Monk, {\em Finite Element Methods for Maxwell's Equations}, Oxford
  University Press, 2003.

\bibitem{Burger2005bacus}
S.~Burger, R.~K\"ohle, L.~Zschiedrich, W.~Gao, F.~Schmidt, R.~M\"arz, and
  C.~N\"olscher, ``Benchmark of {FEM}, {W}aveguide and {FDTD} {A}lgorithms for
  {R}igorous {M}ask {S}imulation,'' in {\em Photomask Technology},  J.~T. Weed
  and P.~M. Martin, eds.,  {\bf 5992}, pp.~378--389, Proc. SPIE, 2005.

\bibitem{Burger2005a}
S.~Burger, R.~Klose, A.~Sch\"adle, and F.~S. and L.~Zschiedrich, ``{FEM}
  modelling of 3{D} photonic crystals and photonic crystal waveguides,'' in
  {\em Integrated Optics: Devices, Materials, and Technologies IX},  Y.~Sidorin
  and C.~A. W\"achter, eds.,  {\bf 5728}, pp.~164--173, Proc. SPIE, 2005.

\bibitem{Enkrich2005a}
C.~Enkrich, M.~Wegener, S.~Linden, S.~Burger, L.~Zschiedrich, F.~Schmidt,
  C.~Zhou, T.~Koschny, and C.~M. Soukoulis, ``Magnetic metamaterials at
  telecommunication and visible frequencies,'' {\em Phys. Rev. Lett.} {\bf 95},
  p.~203901, 2005.

\bibitem{Kalkbrenner2005a}
T.~Kalkbrenner, U.~H{\aa}kanson, A.~Sch\"adle, S.~Burger, C.~Henkel, and
  V.~Sandoghdar, ``Optical microscopy using the spectral modifications of a
  nano-antenna,'' {\em Phys. Rev. Lett.} {\bf 95}, p.~200801, 2005.

\bibitem{Zschiedrich2005b}
L.~Zschiedrich, S.~Burger, A.~Sch\"adle, and F.~Schmidt, ``Domain decomposition
  method for electromagnetic scattering problems,'' in {\em Proceedings of the
  5th International Conference on Numerical Simulation of Optoelectronic
  devices},  pp.~55--56, 2005.

\bibitem{Zsch08}
L.~Zschiedrich, S.~Burger, A.~Sch{\"a}dle, and F.~Schmidt, ``A new domain
  decomposition method for rigorous electromagnetic field simulations,'' in
  {\em Advanced Lithography: Optical Microlithography XXI},   {\bf 6924-193},
  Proc. SPIE, 2008.

\bibitem{Pomplun2007}
J.~Pomplun, S.~Burger, F.~Schmidt, F.~Scholze, C.~Laubis, and U.~Dersch,
  ``Finite {E}lement {A}nalysis of {EUV} {L}ithography,'' in {\em {M}odeling
  {A}spects in {O}ptical {M}etrology},  H.~Bosse, B.~Bodermann, and R.~M.
  Silver, eds.,  {\bf 6617}, p.~18, Proc. SPIE, 2007.

\bibitem{Gross08}
H.~Gross, A.~Rathsfeld, F.~Scholze, R.~Model, and M.~B{\"a}r, ``Computational
  methods estimating uncertainties for profile reconstruction in
  scatterometry,'' in {\em Optical {M}icro- and {N}anometrology in
  {M}icrosystems {T}echnology},   {\bf 6995}, Proc. SPIE, 2008, in press.

\bibitem{Scholze06}
F.~Scholze, C.~Laubis, C.~Buchholz, A.~Fischer, A.~Kampe, S.~Pl{\"o}ger,
  F.~Scholz, and G.~Ulm, ``Polarization dependence of multilayer reflectance in
  the {EUV} spectral range,'' in {\em Emerging Lithographic Technologies XI},
  M.~J. Lercel, ed.,  {\bf 6517}, pp.~863--870, Proc. SPIE, 2006.

\bibitem{Scholze07}
F.~Scholze, C.~Laubis, U.~Dersch, J.~Pomplun, S.~Burger, and F.~Schmidt, ``The
  influence of line edge roughness and {CD} uniformity on {EUV} scatterometry
  for {CD} characterization of {EUV} masks,'' in {\em {M}odeling {A}spects in
  {O}ptical {M}etrology},  H.~Bosse, B.~Bodermann, and R.~M. Silver, eds.,
  {\bf 6617}, p.~1A, Proc. SPIE, 2007.

\end{thebibliography}
\bibliographystyle{spiebib}

\end{document}